\documentstyle[aps,preprint,prb]{revtex}
\def\vk{\vec k} 
\def\vk'{\vec k'} 
\def\hb{\hfill\break}

\def\br{{\bf r}}
\begin{document}

\title{\bf Reinvestigation of Inhomogeneous Superconductors\\
 and High $T_{c}$ Superconductors}
\author{Yong-Jihn Kim }
\address{Department of Physics,  Korea Advanced Institute of Science and 
Technology,\\
 Taejon 305-701, Korea}
\maketitle
\begin{abstract}

We review the recent progress in the theory of inhomogeneous superconductors.
It was shown that Gor'kov's self-consistency equation needs a pairing
constraint derived from the Anomalous Green's function. The Bogoliubov-de Gennes
equations also need a pairing constraint in order to obtain  a correct vacuum
state by the corresponding unitary transformation. 
This new study opens up a reinvestigation of inhomogeneous superconductors.
We discuss (i) problems of the conventional Green's function theory,
(ii) reinvestigation of impure superconductors, and (iii) impurity doping effect
in high $T_{c}$ superconductors.
It is also pointed out that a new formalism is required to tackle the macroscopically or 
mesoscopically inhomogeneous systems such as the junction and the vortex problems.

\vspace{1pc}

\end{abstract}

\vfill\eject

\vskip 1pc
{\bf\Large
\centerline{\bf Reinvestigation of }
\vskip 1pc 
\centerline{\bf Inhomogeneous Superconductors}
\vskip 0.8pc 
\centerline{\bf and }
\vskip 1pc 
\centerline{\bf High $T_{c}$ Superconductors}

\vskip 4pc 

{\bf---------------------------------------------------------------}
\begin{equation}
\Delta({\bf r})=VT\sum_{\omega}\int\{G^{\uparrow}_{\omega}({\bf r},
{\bf l})G^{\downarrow}_{-\omega}({\bf r},{\bf l})\}^{\rm P}\Delta({\bf l})
{\rm d}{\bf l}  
\end{equation}
\centerline{P : Pairing constraint\ \ \ }

{\bf---------------------------------------------------------------}
\vskip 4pc \hb
\rightline{\bf Yong-Jihn Kim\ \ \ \ }
\vskip 1pc \hb
\rightline{ KAIST \quad \quad \ \ \ \ }
}

\vfill\eject
\hspace{2pc}
\centerline{\huge\bf Contents }

\vspace{3pc}
\leftline{\Large\bf Chapter 1 \ Introduction  \quad ................................... 1 }

\vspace{1pc}

\leftline{\large\bf 1. Problems of the conventional Green's function theory  .....1 }

\vspace{1pc}

 1.1. Remarks on the failure of Green's function theory ..................................... 2

 1.2. Discrepancy between Pippard theory and Ginzburg-Landau theory ........ 12

 1.3. Correspondence principle and theory of superconductivity ...................... 14

 1.4. Pairing constraint on Gor'kov formalism .................................................. 19

 1.5. Pairing constraint on the Bogoliubov-de Gennes equations ...................... 20

 1.6. Relation between pair potential and gap parameter ................................. 23

 1.7. Problem in the microscopic derivation of the Ginzburg-Landau 

\hspace{2pc} theory ...................................................................................................... 24

\vspace{1pc}

\leftline{\large\bf 2. Reinvestigation of impure superconductors \quad ............... 25}

\vspace{1pc}

 2.1. History of the theory of impure superconductors ...................................... 26

 2.2. New theory of impure superconductors by Kim and Overhauser .............. 27

 2.3. Strong-coupling theory of impure superconductors ................................... 30

 2.4. Compensation of magnetic impurity effect by radiation damage 

\hspace{1pc} or ordinary impurity ................................................................................... 33

 2.5. Localization and superconductivity .......................................................... 35

\vspace{1pc}

\leftline{\large\bf 3. Study of high Tc superconductors \quad ............................. 38}

\vspace{1pc}

 3.1. Impurity scattering in a d-wave superconductor ........................................ 39

 3.2. Impurity doping effect in high Tc  superconductors ................................... 42

 3.3. On the mechanism of high Tc superconductors ....................................... 44

 3.4. Search for new high Tc superconductors ................................................... 45

\vspace{1pc}

\leftline{\large\bf 4. New formalism for inhomogeneous superconductors \ \  .. 46}

\vspace{1pc}

 4.1. Generalization of the BCS theory ............................................................. 47

 4.2. Generalization of the Ginzburg-Landau theory ........................................ 47

\vspace{1pc}

\leftline{\large\bf 5. Future directions \quad ....................................................... 48}

\vspace{1pc}

\hspace{1pc} 5.1. Impure superconductors ...................................................................... 49

\hspace{1pc} 5.2. Localization and Superconductivity .................................................... 49

\hspace{1pc} 5.3. Proximity effect ................................................................................... 50

\hspace{1pc} 5.4. Andreev reflection ................................................................................ 51

\hspace{1pc} 5.5. Josephson effect ................................................................................... 51

\hspace{1pc} 5.6. Magnetic field effect ............................................................................. 51

\hspace{1pc} 5.7. Type II superconductors ...................................................................... 54

\hspace{1pc} 5.8. Vortex problem .................................................................................... 54

\hspace{1pc} 5.9. Mesoscopic superconductivity .............................................................. 55

\hspace{1pc} 5.10. Non-equilibrium superconductivity .................................................... 56

\hspace{1pc} 5.11. Granular superconductors .................................................................. 56

\hspace{1pc} 5.12. High $T_{c}$ superconductors ................................................................... 56

\vspace{1pc}

\leftline{\Large\bf Chapter 2 \ Basic Formalism \quad ........................... 57}

\vspace{1pc}

\noindent
{\bf 1. ``A constraint on the Anomalous Green's function"}, 

{\sl Yong-Jihn Kim},

Mod. Phys. Lett. B{\bf 10}, 555 (1996). ................................................................ 58

\noindent
{\bf 2. ``Pairing in the Bogoliubov-de Gennes equations"}, 

{\sl Yong-Jihn Kim},

Int. J.  Mod. Phys. B{\bf 11}, 1731 (1997). ............................................................ 69

\vspace{1pc}

\leftline{\Large\bf Chapter 3  Theory of impure superconductors \ .. 81}
\vspace{1pc}

\noindent
{\bf 3. ``Theory of impure superconductors: Anderson versus Abrikosov}
 
\hspace{1pc} {\bf and Gor'kov"}, {\sl Yong-Jihn Kim and A. W. Overhauser},

Phys. Rev. B {\bf 47}, 8025 (1993). ........................................................................ 82

\noindent
{\bf 4. ``Comment on `Theory of impure superconductors: Anderson} 

\hspace{1pc} {\bf versus Abrikosov and Gor'kov'} ", 

{\sl A. A. Abrikosov and L. P. Gor'kov},

Phys. Rev. B{\bf 49}, 12339 (1994). ....................................................................... 87

\noindent
{\bf 5. Reply to ``Comment on `Theory of impure superconductors:} 

\hspace{1pc} {\bf Anderson versus Abrikosov and Gor'kov'} ", 

{\sl Yong-Jihn Kim and A. W. Overhauser},

Phys. Rev. B{\bf 49}, 12339 (1994). ....................................................................... 89 

$\bullet$ Communication from A. A. Abrikosov. ........................................................ 91

$\bullet$ P. 337 in {\sl Methods of Quantum Field Theory in Statistical Physics. ............. 92}

$\bullet$ Communication from D. M. Ginsberg. ......................................................... 93     

\noindent
{\bf 6. ``Magnetic impurities in superconductors:
A theory with }

\hspace{1pc} {\bf different predictions"}, {\sl Yong-Jihn Kim and A. W. Overhauser}, 

Phys. Rev. B{\bf 49}, 15779 (1994). ....................................................................... 94

$\bullet$ Communication from P. W. Anderson. ....................................................... 108 

$\bullet$ P. W. Anderson's rapporteur's remarks at Toronto LTIV. ......................... 109 

$\bullet$ Comment by Muthu and Bhardwaj. ........................................................... 110 

\noindent
{\bf 7. ``Strong-coupling theory of impure superconductors"}, 

{\sl Yong-Jihn Kim},  

Mod. Phys. Lett. B, {\bf 10}, 353-358 (1996). ...................................................... 111

\noindent
{\bf 8. ``Weak localization effect in supercondutors"}, 

{\sl Yong-Jihn Kim and K. J. Chang},

submitted to Phys. Rev. B (1997). ............................................................... 117

\noindent
{\bf 9. ``Compensation of magnetic impurity effect in superconductors}

\hspace{1pc}{\bf by radiation damage}" 

{\sl M.-A. Park, M. H. Lee and Yong-Jihn Kim}, 

submitted to Physica C (1997). .................................................................... 136

\noindent
{\bf 10. ``Strong coupling theory of impure superconductors: }

\hspace{1pc} {\bf Correspondence with weak coupling theory}", 

{\sl Yong-Jihn Kim}, 

submitted to Phys. Rev. B (1997). ............................................................... 148

\vspace{1pc}

\leftline{\Large\bf Chapter 4 \ Short Reviews \quad ............................. 164}
\vspace{1pc}

\noindent
{\bf 11. ``Pairing constraint on the real space formalism of the theory of}

\hspace{1pc} {\bf  superconductivity"}, {\sl Yong-Jihn Kim}, 

to appear in Proceedings  of Inauguration Conference of Asia Pacific 

Center for Theoretical Physics, World Scientific, 1997. .................................. 165

\noindent
{\bf 12. ``Impurity scattering in supercondutors}", 

{\sl Yong-Jihn Kim and K. J. Chang},

J. Korean Phys. Soc. Vol. {\bf 31}, S298-S301 (1997). ......................................... 179

\vspace{1pc}

\leftline{\Large\bf Chapter 5 \ High Tc superconductors \quad ............ 183}

\vspace{1pc}

\noindent
{\bf 13. ``Impurity scattering in a d-wave supercondutor"}, 

{\sl M.-A. Park, M. H. Lee and  Yong-Jihn Kim}, 

Mod. Phys. Lett. B11, 16 \& 17, 719 (1997). .................................................. 184

\noindent
{\bf 14. ``Impurity doping effect in high Tc in supercondutors"}, 

{\sl Yong-Jihn Kim and K. J. Chang}, 

submitted to Phys. Rev. B (1997). ................................................................ 192

\vspace{1pc}

\leftline{\Large\bf Chapter 6 \ Superfluid He-3 \quad ........................... 205}
\vspace{1pc}

\noindent
{\bf 15. ``Extra triplet-pairing states for superfluid $^{3}$He"}, 

{\sl Yong-Jihn Kim and A. W. Overhauser}, 

preprint. ........................................................................................................ 206

\vfill\eject

\hspace{2pc}
\centerline{\huge\bf Chapter 1 \ Introduction }

\vspace{4pc}

\leftline{\large\bf 1. Problems of the conventional Green's function theory }

\vspace{2pc}

 1.1. Remarks on the failure of Green's function theory 

 1.2. Discrepancy between Pippard theory and Ginzburg-Landau theory

 1.3. Correspondence principle and theory of superconductivity

 1.4. Pairing constraint on Gor'kov formalism

 1.5. Pairing constraint on the Bogoliubov-de Gennes equations

 1.6. Relation between pair potential and gap parameter

 1.7. Problem in the microscopic derivation of the Ginzburg-Landau 

\hspace{2pc} theory

\hspace{5pc}

\hspace{2pc}
\noindent

\centerline{\large\bf 1.1. Remarks on the failure of Green's function theory} 
\vspace{1pc}

{\bf Impure superconductors}

\vspace{1pc}

$\bullet$ ``Theory of superconductivity", by {\sl P. W. Anderson}, rapporteur's
remarks 

at Toronto LTIV, 1960.

\vspace{1pc}

By the use of Gor'kov techniques, Abrikosov and Gor'kov have succeeded in obtaining
a perturbation theory valid for very small energy gap $\cdots$, 
it is entirely \underbar{incorrect} as far as any physical results are concerned.

\vspace{1pc}

$\bullet$ ``Breakdown of Eliashberg theory for two-dimensional superconductivity in

the presence of disorder" by {\sl R. C. Dynes, A. E. White, J. M. Graybeal,}

{\sl  and J. P. Garno}, Phys. Rev. Lett. {\bf 57}, 2195 (1986).

\vspace{1pc}

Abstract: We conclude that the standard picture of enhanced Coulomb repulsion with 
increasing sheet resistance is \underbar{too naive} and that, in the presence of disorder,
an analysis more sophisticated than Eliashberg theory is necessary. 

\vspace{1pc}

$\bullet$ ``Apparent destruction of superconductivity in the
disordered one-dimensional

limit" by {\sl J. M. Graybeal, P. M. Mankiewich, R. C. Dynes, and M. R. Beasley}, 

Phys. Rev. Lett. {\bf 59}, 2697 (1987).

\vspace{1pc}

Abstract: Our findings are in \underbar{clear dsagreement} with a recent theoretical treatment.

$\bullet$ ``Superconducting-Insulating transition in two-dimensional
a-MoGe thin films

by {\sl A. Yazdani and A. Kapitulnik}, Phys. Rev. Lett. {\bf 74}, 3037 (1995).

\vspace{1pc}

Abstract: However, \underbar{contrary to} the theoretical predictions we find the critical
resistance to be sample dependent.

\vspace{1pc}

{\bf Junction Problems}

\vspace{1pc}

$\bullet$ ``Boundary-condition effects on the superconducting transition temperature

of proximity-effect systems", by {\sl P. R. Broussard}, Phys. Rev. B {\bf 43}, 2783 (1991).

\vspace{1pc}

Abstract: The superconucting critical temperature, $T_{c}$, of different configurations of
layers is studied under the de Gennes-Werthamer model. \underbar{Certain Inconsistencies}
are seen to develop with the use of this approach, calling into question previous
results obtained.

\vspace{1pc}

$\bullet$ ``Introduction to superconductivity", by {\sl M. Tinkham}, McGraw-Hill, 1975.

\vspace{1pc}

P. 195.

\vspace{1pc}

Thus, for $V\not= 0$, and for an arbitray type of weak-link element, we may
generalize (6-4) to

\hspace{5pc} $I=I_{o}sin\gamma + (G_{o}+G_{int} cos\gamma )V$ \hspace{8pc} (6-4a)
\hfill\break
where $G_{o}$, $G_{int}$ themselves may be functions of V.
Experiments of Pedersen et al.$^{1}$ on tunnel junctions, of Falco et al.$^{2}$
on thin-film weak links, and of Vincent and Deaver$^{3}$ on point-contact weak links have all demonstrated the existence of this pair-quasi-particle
interference term, and all have shown that $G_{int}/G_{o} \approx \underbar{-1}$. 
This result has drawn much recent attention, because it appears that 
the microscopic theory$^{4}$ yields a \underbar{positive sign} for this ratio. 

\vspace{1pc}

$\bullet$ ``Superconducting weak links", by {\sl K. K. Likharev}, Rev. Mod. Phys. {\bf 51}, 101 

(1979).

\vspace{1pc}

B. Unsolved problems

2. ac processes

(1) Is the discrepancy in the signs of the same two terms between experiment
and the Mitsai theory indicative of certain \underbar{fundamental inconsistencies}
in the Green's function formulation of the theory of nonstationary
superconductivity?

C. Concluding remarks

$\cdots$ in the field of ac effects there are more questions than answers.

\vspace{2pc}

\vspace{2pc}
\noindent

\leftline{\large\bf 1.2. Discrepancy between Pippard theory and Ginzburg-}
\centerline{\large\bf  Landau theory} 
\vspace{1pc}

{\bf Pippard-BCS theory}

\vspace{1pc}

Pippard assumed that the current ${\bf j}({\bf r})$ at one point will depend
on the vector potential ${\bf A}({\bf r}')$ at all neighboring
points ${\bf r}'$ such that $|{\bf r}-{\bf r}'|<\xi_{o}$.
Here $\xi_{o}$ means the BCS coherence length
\begin{equation}
\xi_{o}= {\hbar v_{F}\over \pi \Delta}.
\end{equation}

The Pippard nonlocal current relation becomes

\begin{equation}
{\bf j}({\bf r})=-{3\over 4\pi\xi_{o}c\Lambda}\int {{\bf R}[{\bf R}\cdot {\bf E}
({\bf r}')]\over R^{4}}e^{-R/\xi_{o}}d^{3}r',\hspace{3pc}
{\bf R}={\bf r}-{\bf r}'.
\end{equation}

Note that the range of nonlocality is reduced by a factor of about 0.75 on
going from $T=0$ to $T_{c}$.

It is a tribute to Pippard's insight into the physics of superconductivity
that his equation is almost identical to that given by BCS theory.

\vspace{1pc}

{\bf Ginzburg-Landau theory}

\vspace{1pc}

The fundamental Landau-Ginzburg equations are the following:

\begin{eqnarray}
\alpha \psi &+& \beta|\psi|^{2}\psi + {1\over 2m}(-i\hbar\nabla -{2e{\bf A}\over c})^{2}\psi = 0, \\
{\bf j} &=& {e\hbar\over im}(\psi^{*}\nabla \psi - \psi\nabla \psi^{*})
- {4e^{2}\over mc}\psi^{*}\psi{\bf A}.
\end{eqnarray}

\vspace{1pc}

Even though the relation between ${\bf j}$ and $\bf A$ is approximated by a 
local relation, the Ginzburg-Landau theory definitely includes nonlocal 
effects and the coherence length appears in a natural way.
The Ginzburg-Landau coherence length $\xi_{GL}$ is given by
\begin{equation}
\xi_{GL}= 0.74{\xi_{o}\over \sqrt{1-T/T_{c}}}.
\end{equation}
(From J. R. Schrieffer, Theory of Superconductivity, p. 22, Benjamin, 1964.)

However, the Ginzburg-Landau coherence length $\xi_{GL}$ is very different 
from the BCS coherence length $\xi_{o}$.
(From Fetter and Walecka, Quantum theory of many particle systems, p. 433, 1971.)

Note that the range of nonlocality in the Pippard-BCS theory is reduced by a 
factor of about 0.75 on going from $T=0$ to $T_{c}$.
Since the Pippard-BCS theory does not allow the spatial variation of the 
energy gap or the order parameter $\psi({\bf r})$,
the nonlocal effect in the Pippard-BCS theory is different from 
that in the Ginzburg-Landau theory.
In other words, we should compare the Pippard nonlocal relation Eq. (3) with 
the Ginzburg-Landau local relation Eq. (5) for a spatially uniform
order parameter $\psi({\bf r})$.

{\bf In conclusion, the Ginzburg-Landau theory is in serious conflict with the
Pippard-BCS theory.}

\vspace{2pc}

\hspace{2pc}
\noindent

\leftline{\large\bf 1.3. Correspondence principle and theory of superconductivity}

\vspace{2pc}

{\bf 1.3.1. Abstract for 1998 Theoretical Solid State Physics Symposium}

 Taejon, Feb. 19-21, 1998, Korea

\vspace{1pc}

\centerline{\bf Correspondence Principle and Theory of Superconductivity$^{\dagger}$ }

\vspace{1pc}

\centerline{Yong-Jihn Kim }

\centerline{Department of Physics,  Korea Advanced Institute of Science and 
Technology}

\centerline{ Taejon 305-701, Korea}

\vspace{1pc}

\centerline{Abstract}

The conventional theory of superconductivity has a hierarchical structure.
After the microscopic theory by Bardeen, Cooper, and Schrieffer,  there 
appeared Gor'kov formalism, the Bogoliubov-de Gennes equations, and the 
Eliashberg formalism which tried to generalize the BCS theory. On the 
macroscopic level, the London theory, the Pippard's nonlocal
theory, and the Ginzurg-Landau theory form such a hierarchical structure. 
We discuss the correspondence between the theories. It is shown that the 
correspondence principle leads to the microscopic pairing constraint and 
therefore the theory of inhomogeneous superconductors needs to be reinvestigated.   
In particular, since the Ginzburg-Landau theory lacks the information of 
the pairing correlations it gives rise to different nonlocal electrodynamics 
from those of the Pippard and the BCS theories.

\vspace{1pc}
\noindent
$\dagger$ To the memory of Professor D. J. Kim.

\vspace{1pc}

{\bf 1.3.2. Correspondence between BCS theory and Gor'kov formalism}

\vspace{1pc}

{\bf BCS Theory}

\vspace{1pc}

For a homogeneous system,  
\begin{eqnarray}
H_{red} = \sum_{{\vec k} {\vec k}'} V_{{\vec k} {\vec k}'} c_{{\vec k}'}^{\dagger}c_{-{\vec k}'}^{\dagger}
c_{-{\vec k}}c_{{\vec k}}, 
\end{eqnarray}
where
\begin{eqnarray}
V_{{\vec k}{\vec k}'}= \cases{-V, &if $|\epsilon_{{\vec k}}|,|\epsilon_{{\vec k}'}|\leq  \omega_{D}$\cr
                      0, &otherwise.\cr}
\end{eqnarray}
This reduction procedure is recognizing in advance which eigenstates
will be paired and so contribute to the BCS condensate.

The BCS gap equation:
\begin{eqnarray}
\Delta_{\vec k}=\sum_{n'}V_{{\vec k}{\vec k}'}{\Delta_{\vec k'}\over 2E_{\vec k'}}tanh
{E_{\vec k'}\over 2T}.
\end{eqnarray}

\vspace{1pc}

{\bf Gor'kov formalism}

\vspace{1pc}

In Gor'kov formalism, a point
interaction $-V\delta({\bf r}_{1}-{\bf r}_{2})$ is used for the 
pairing interaction between electrons. 
For a homogeneous system, the pairing interaction is
\begin{eqnarray}
H_{G} &=& - {1\over 2}V\int d{\bf r}\sum_{\alpha\beta}\Psi^{\dagger}({\bf r}\alpha)
\Psi^{\dagger}({\bf r}\beta)\Psi({\bf r}\beta)\Psi({\bf r}\alpha) \nonumber \\
&=&-{1\over 2}V\sum_{{\vec k}{\vec k}'{\vec q}\sigma\sigma '}
c_{{\vec k} - {\vec q}, \sigma}^{\dagger}c_{{\vec k}' + {\vec q}, \sigma '}^{\dagger}c_{{\vec k} '\sigma '}
c_{{\vec k}, \sigma}, 
\end{eqnarray}
and
\begin{eqnarray}
V_{{\vec k}{\vec k}'}&=&-V\int \phi_{{\vec k}'}^{*}({\bf r}) \phi_{-{\vec k}'}^{*}({\bf r})
 \phi_{-{\vec k}}({\bf r}) \phi_{{\vec k}}({\bf r})d{\bf r}\nonumber \\
&=&-V.
\end{eqnarray}
The self-consistency equation is
\begin{eqnarray}
\Delta({\bf r}) = VT\sum_{\omega}\int \Delta({{\bf l}})G^{\uparrow}_{\omega}({\bf r,{\bf l}})
G^{\downarrow}_{-\omega}({\bf r,{\bf l}})d{\bf l}.
\end{eqnarray}

\vspace{1pc}

{\bf Correspondence principle}

\vspace{1pc}

Note that the two points are not clear in Gor'kov's formalism, i.e.,
the BCS reduction procedure and the retardation cutoff.
To obtain the same result as that of the BCS theory,
these two ingredients should be taken care of in some way.
As will be shown later, the negligence of the BCS reduction procedure
causes a serious pairing problem especially in impure superconductors. 

The correspondence principle leads to the revised  self-consistency equation,
\begin{eqnarray}
\Delta({\bf r}) = VT\sum_{\omega}\int \Delta({{\bf l}})\{G^{\uparrow}_{\omega}({\bf r,{\bf l}})
G^{\downarrow}_{-\omega}({\bf r,{\bf l}})\}^{P}d{\bf l},
\end{eqnarray}
where P denotes a pairing constraint, which is derived from the physical constrant of the
Anomalous Green's function,
\begin{equation}
F({\bf r},{\bf r'}) = F({\bf r}-{\bf r'}).
\end{equation}
Notice that Eq. (13) is nothing but another form of the
BCS gap equation.

\vspace{1pc}

{\bf 1.3.3. Correspondence between BCS theory and Eliashberg formalism}

\vspace{1pc}

{\bf Eliashberg theory with pairing constraint} 

The conventional self-consistency equation for the pair potential
is
\begin{eqnarray}
\Delta^{*}(&\omega_{n}&, {\bf r})Z(\omega_{n}) \nonumber\\
 &=& \gamma^{2}  T\sum_{n'}\lambda(\omega_{n},\omega_{n'})\int d{\bf r}_{o}
G_{N}^{\uparrow}(-\omega_{n'},{\bf r}_{o},
{\bf r})G_{N}^{\downarrow}(\omega_{n'},{\bf r}_{o},{\bf r})\Delta^{*}(\omega_{n'},{\bf r}_{o})
Z(\omega_{n'}).  
\end{eqnarray}
From the physical constraint of the Anomalous Green's function, i.e.,
\begin{eqnarray}
F^{+}(\omega_{n},{\bf r},{\bf r'})= 
F^{+}(\omega_{n},{\bf r}-{\bf r'}), 
\end{eqnarray}
we find the revised self-consistency equation,
\begin{eqnarray}
\Delta^{*}(&\omega_{n}&, {\bf r})Z(\omega_{n})  = \nonumber\\
 & \gamma^{2} & T\sum_{n'}\lambda(\omega_{n},\omega_{n'})\int d{\bf r}_{o}
\{G_{N}^{\uparrow}(-\omega_{n'},{\bf r}_{o}, {\bf r})
G_{N}^{\downarrow}(\omega_{n'},{\bf r}_{o},{\bf r})\}^{P}
\Delta^{*}(\omega_{n'},{\bf r}_{o})Z(\omega_{n'}). 
\end{eqnarray}
Since
\begin{equation}
\Delta^{*}(\omega_{n}, m) = \int \psi_{m}({\bf r})\psi^{*}_{m}({\bf r})
\Delta^{*}(\omega_{n},{\bf r})d{\bf r},
\end{equation}
the strong-coupling gap equation is
\begin{equation}
\Delta^{*}(\omega_{n}, m)Z(\omega_{n}) = T
\sum_{n'}\lambda(\omega_{n},\omega_{n'})
 \sum_{m'}V_{mm'}{\Delta^{*}(\omega_{n'},m')Z(\omega_{n'})\over [\omega_{n'}Z(\omega_{n'})]^{2}
+\epsilon_{m'}^{2}},
\end{equation}
where
\begin{equation}
V_{mm'} = \gamma^{2}\int |\psi_{m}({\bf r})|^{2}
 |\psi_{m'}({\bf r})|^{2}d{\bf r}.
\end{equation}

{\bf Weak-coupling limit} 

The strong-coupling theory leads to the weak-coupling theory in the static limit, (i.e.),
\begin{eqnarray}
\Delta^{*}(\omega_{n},{\bf r})&=&\Delta^{*}(0, {\bf r})=\Delta^{*}({\bf r}),\\
Z(\omega)&=&Z(0)=1,\\
\lambda(\omega_{n},\omega_{n'})&=&\lambda(0,0)=1.
\end{eqnarray}
Accordingly, we find
\begin{equation}
\Delta^{*}( {\bf r})= \gamma^{2}  T\sum_{n'}\int d{\bf r}_{o}
G_{N}^{\uparrow}(-\omega_{n'},{\bf r}_{o},
{\bf r})G_{N}^{\downarrow}(\omega_{n'},{\bf r}_{o},{\bf r})\Delta^{*}({\bf r}_{o}) ,  
\end{equation}
\begin{equation}
\Delta^{*}({\bf r}) =  \gamma^{2} T\sum_{n'}\int d{\bf r}_{o}
\{G_{N}^{\uparrow}(-\omega_{n'},{\bf r}_{o}, {\bf r})
G_{N}^{\downarrow}(\omega_{n'},{\bf r}_{o},{\bf r})\}^{P}
\Delta^{*}({\bf r}_{o}), 
\end{equation}
and
\begin{eqnarray}
\Delta^{*}(m) &=& T
\sum_{n'} \sum_{m'}V_{mm'}{\Delta^{*}(m')\over \omega_{n'}^{2}
+\epsilon_{m'}^{2}}\nonumber\\
&=& \sum_{m'}V_{mm'}{\Delta^{*}(m') \over 2\epsilon_{m'}} 
tanh({\epsilon_{m'}\over 2T}).
\end{eqnarray}

{\bf Correspondence}

Note that the correspondence principle, which relates strong-coupling and weak-coupling
theories, works only when the pairing constraint is incorporated into the self-consistency
equation. For more details, see Y.-J. Kim, ``Strong-coupling theory of impure superconductors:
Correspondence with weak-coupling theory". 

\vspace{1pc}

{\bf 1.3.4. Correspondence between Pippard-BCS theory and Ginzburg-}

\hspace{9pc} {\bf Landau theory}

\vspace{1pc}

Superconductivity is a typical example of macroscopic quantum phenomena which 
can be explained only by the quantum mechanics. 
Since the Ginzburg-Landau theory was constructed in 1950 before the pairing theory of
BCS, it lacks the microscopic information of the pairing correlation and  
leads to different nonlocal electrodynamics from that of the Pippard-BCS theory.
(See also Ch. 1.2.)
Gor'kov's microscopic derivation of the Ginzburg-Landau theory is not valid because
he used the incorrect self-consistency equation.

We should generalize the Ginzburg-Landau theory by including  the information of the
pairing correlation.

\vspace{2pc}

\hspace{2pc}
\noindent

\leftline{\large\bf 1.4. Pairing constraint on Gor'kov formalism}

\vspace{1pc}

From Y.-J. Kim, ``A constraint on the Anomalous Green'sfunction", Mod. Phys. 

\hspace{3pc} Lett. B, (1996).

\vspace{1pc}

{\bf Homogeneous System}

The conventional self-consistency equation is
\begin{eqnarray}
\Delta({\bf r}) = VT\sum_{\omega}\int \Delta({{\bf l}})G^{\uparrow}_{\omega}({\bf r,{\bf l}})
G^{\downarrow}_{-\omega}({\bf r,{\bf l}})d{\bf l}.
\end{eqnarray}
From the physical constraint of the Anomalous Green's function,
\begin{equation}
F({\bf r},{\bf r'}) = F({\bf r}-{\bf r'}),
\end{equation}
we obtain the revised self-consistency equation,
\begin{eqnarray}
\Delta({\bf r}) = VT\sum_{\omega}\int \Delta({{\bf l}})\{G^{\uparrow}_{\omega}({\bf r,{\bf l}})
G^{\downarrow}_{-\omega}({\bf r,{\bf l}})\}^{P}d{\bf l},
\end{eqnarray}
where P denotes a pairing constraint which dictates pairing between ${\vec k}\uparrow$
and ${\vec k}\downarrow$. 

Accordingly, the revised strong-coupling self-consistency equation is
\begin{eqnarray}
\Delta^{*}(&\omega_{n}&, {\bf r})Z(\omega_{n})  = \nonumber\\
 & \gamma^{2} & T\sum_{n'}\lambda(\omega_{n},\omega_{n'})\int d{\bf r}_{o}
\{G_{N}^{\uparrow}(-\omega_{n'},{\bf r}_{o}, {\bf r})
G_{N}^{\downarrow}(\omega_{n'},{\bf r}_{o},{\bf r})\}^{P}
\Delta^{*}(\omega_{n'},{\bf r}_{o})Z(\omega_{n'}). 
\end{eqnarray}

\vspace{1pc}

{\bf Dirty System}

From the physical constraint of the Anomalous Green's function, i.e.,
\begin{eqnarray}
\overline{F({\bf r},{\bf r'},\omega)}^{imp} 
= \overline{F({\bf r}-{\bf r'},\omega)}^{imp},
\end{eqnarray}
the revised  self-consistency equation is
\begin{eqnarray}
\Delta({\bf r}) = VT\sum_{\omega}\int \Delta({{\bf l}})\{G^{\uparrow}_{\omega}({\bf r,{\bf l}})
G^{\downarrow}_{-\omega}({\bf r,{\bf l}})\}^{P}d{\bf l},
\end{eqnarray}
where P denotes the Anderson's pairing condition.
Notice that Eq. (32) is nothing but another form of the
BCS gap equation,
\begin{eqnarray}
\Delta_{n}=\sum_{n'}V_{nn'}{\Delta_{n'}\over 2E_{n'}}tanh
{E_{n'}\over 2T}.
\end{eqnarray}

\vspace{2pc}

\hspace{2pc}
\noindent

\leftline{\large\bf 1.5. Pairing constraint on the Bogoliubov-de Gennes equations}

\vspace{1pc}

From Y.-J. Kim, ``Pairing in the Bogoliubov-de Gennes equations", Int. J. Mod. 

\hspace{3pc} Phys. Lett. B, (1997).

\vspace{1pc}

{\bf 1.5.1. Dirty System}

In 1959, P. W. Anderson introduced a ground state for dirty superconductors,
which is given by
\begin{eqnarray}
{\tilde \phi_{Anderson}} = \prod_{n} (u_{n} + v_{n}c_{n\uparrow}^{\dagger}
         c_{{\bar n}\downarrow}^{\dagger})|0>, 
\end{eqnarray}
where $c_{{\bar n}\downarrow}^{\dagger}$ is the creation operator for an electron
in the scattered-state $\psi_{n}^{*}({\bf r})|\downarrow>$.
On the other hand, it has been claimed that the energy is lowered
if we pair states $ \Phi_{n}$ which are better choices than $\psi_{n}$ 
by using the Bogoliubov-de Gennes equations. 
The state $ \Phi_{n}$ is basically a linear combination of the
normal scattered states.
The coupling comes from the pair potential.
However, we show that 
pairing $ \Phi_{n}\uparrow$ and $ \Phi_{\bar n}\downarrow$
leads to the violation of the physical constraint of the system.

The unitary transformation,
\begin{eqnarray}
\Psi({\bf r}\uparrow) &=& \sum_{n}(\gamma_{n\uparrow}u_{n}({\bf r}) - 
\gamma^{\dagger}_{n\downarrow}v^{*}_{n}({\bf r})), \nonumber\\
\Psi({\bf r}\downarrow) &=& \sum_{n}(\gamma_{n\downarrow}u_{n}({\bf r}) + 
\gamma^{\dagger}_{n\uparrow}v^{*}_{n}({\bf r})), 
\end{eqnarray}
leads to the following Bogoliubov-de Gennes equations,
\begin{eqnarray}
\epsilon u({\bf r}) &=& [H_{e}+U({\bf r})]u({\bf r}) + \Delta({\bf r})v({\bf r}), \nonumber\\ 
\epsilon v({\bf r}) &=& -[H_{e}^{*}+U({\bf r})]v({\bf r}) + \Delta^{*}({\bf r})u({\bf r}). 
\end{eqnarray}
To find the vacuum state for $\gamma$ particles, we expand the field
operator by the scattered states:
\begin{eqnarray}
\Psi({\bf r}\alpha) = \sum_{n} \psi_{n}({\bf r}) c_{n\alpha}.
\end{eqnarray}
Then it can be shown 
\begin{eqnarray}
\gamma_{n\uparrow} &=&\sum_{n'}\bigl( u_{n,n'}^{*}c_{n'\uparrow} + v_{n,n'}c_{n'\downarrow}^{\dagger}\bigr)=U_{n}b_{n\uparrow}-V_{n}b^{\dagger}_{{\bar n}\downarrow},\nonumber\\
\gamma_{n\downarrow} &=&\sum_{n'}\bigl( u_{n,n'}^{*}c_{n'\downarrow} - v_{n,n'}c_{n'\uparrow}^{\dagger}\bigr)=U_{n}b_{n\downarrow}+V_{n}b_{{\bar n}\uparrow}^{\dagger},
\end{eqnarray}
where
\begin{eqnarray}
u_{n,n'} &=& \int \psi^{*}_{n'}({\bf r}) u_{n}({\bf r})d{\bf r},\nonumber\\
v_{n,n'} &=& \int \psi^{*}_{n'}({\bf r})v^{*}_{n}({\bf r})d{\bf r}. 
\end{eqnarray}
Finally, we obtain 
\begin{eqnarray}
{\tilde \phi_{BdG}} = \prod_{n} (U_{n} + V_{n}b_{n\uparrow}^{\dagger}
         b_{{\bar n}\downarrow}^{\dagger})|0>. 
\end{eqnarray}
Note that the Bogoliubov-de Gennes equations, Eq.(36) correspond to the 
vacuum state where ${\tilde \Phi}_{n}({\bf r})[={1\over U_{n}}u_{n}({\bf r})]\uparrow$ 
and
 ${\tilde \Phi}_{\bar n}({\bf r})[=-{1\over V_{n}}v^{*}_{n}({\bf r})]\downarrow$ (instead of $\psi_{n}({\bf r})\uparrow$
and $\psi_{\bar n}({\bf r})\downarrow$) are paired.

Now we must decide which is the correct ground state in the presence of 
impurities.
Above all, the correct ground state should satisfy the
physical constraint of the system.
It can be shown that the state ${\tilde \phi}_{BdG}$ gives the `averaged' pair potential
\begin{eqnarray}
\overline{\Delta({\bf r})}^{imp} \not= \rm constant, 
\end{eqnarray}
which violates the physical constraint of the system.
Therefore the correct ground state is  
${\tilde \Phi}_{Anderson}$. 
To obtain 
${\tilde \Phi}_{Anderson}$ 
from the Bogoliubov-de Gennes equations,
we need a pairing constraint:
\begin{eqnarray}
{\tilde \phi_{BdG}} = {\tilde \phi_{Anderson}},
\end{eqnarray}
which gives
\begin{eqnarray}
u_{n}({\bf r}) &\propto& \psi_{n}({\bf r}),\nonumber\\
v_{n}^{*}({\bf r}) &\propto& \psi_{\bar n}({\bf r}).
\end{eqnarray}
\vskip 1pc

{\bf 1.5.2. Homogeneous System}

The unitary transformation, 
\begin{eqnarray}
\Psi({\bf r}\uparrow) &=& \sum_{n}(\gamma_{n\uparrow}u_{n}({\bf r}) - 
\gamma^{\dagger}_{n\downarrow}v^{*}_{n}({\bf r})), \nonumber\\
\Psi({\bf r}\downarrow) &=& \sum_{n}(\gamma_{n\downarrow}u_{n}({\bf r}) + 
\gamma^{\dagger}_{n\uparrow}v^{*}_{n}({\bf r})), 
\end{eqnarray}
leads to
\begin{eqnarray}
\gamma_{n'\uparrow} &=& \sum_{\vec {k}}\bigl( u_{n',\vec {k}}^{*}a_{\vec {k}\uparrow} + v_{n',\vec {k}}a_{\vec {k}\downarrow}^{\dagger}\bigr)
= U_{n}a_{n\uparrow} - V_{n}a_{{\bar n}\downarrow}^{\dagger}, \nonumber\\
\gamma_{n'\downarrow} &=& \sum_{\vec {k}}\bigl( u_{n',\vec {k}}^{*}a_{\vec {k}\downarrow} - v_{n',\vec {k}}a_{\vec {k}\uparrow}^{\dagger}\bigr)
= U_{n}a_{n\downarrow} + V_{n}a_{{\bar n}\uparrow}^{\dagger},
\end{eqnarray}
where
\begin{eqnarray}
u_{n,\vec {k}} &=& \int \phi^{*}_{\vec {k}}({\bf r}) u_{n}({\bf r})d{\bf r},\nonumber\\
v_{n,\vec {k}} &=& \int \phi^{*}_{\vec {k}}({\bf r})v^{*}_{n}({\bf r})d{\bf r}.
\end{eqnarray}

Note that we pair $\Phi_{n}({\bf r})[={1\over U_{n}}u_{n}({\bf r})]\uparrow$ and 
$\Phi_{\bar n}({\bf r})[=-{1\over V_{n}}v^{*}_{n}({\bf r})]\downarrow$
(instead of $\phi_{\vec {k}}({\bf r})\uparrow$ and $\phi_{-\vec {k}}({\bf r})\downarrow$)
by the unitary transformation (11). 
The generated vacuum state is
\begin{eqnarray}
{\tilde \phi_{BdG}} = \prod_{n} (U_{n} + V_{n}a_{n\uparrow}^{\dagger}
         a_{{\bar n}\downarrow}^{\dagger})|0>, 
\end{eqnarray}
instead of the BCS ground state
\begin{eqnarray}
{\tilde \phi_{BCS}} = \prod_{\vec {k}} (u_{\vec {k}} + v_{\vec {k}}a_{\vec {k}\uparrow}^{\dagger}
         a_{-\vec {k}\downarrow}^{\dagger})|0>. 
\end{eqnarray}
Therefore a pairing constraint is necessary for the unitary
transformation (11) to generate the BCS ground state; that is, 
both $u_{n}({\bf r})$ and $v_{n}({\bf r})$ should be proportional to the normal
state wavefunction $\phi_{\vec {k}}(\bf r)$ in order to pair $\vec {k}\uparrow$ and
$-\vec {k}\downarrow$.
For the current-carrying state, we can pair $\vec {k}+{\vec q}\uparrow$ and $-\vec {k}+{\vec q}\downarrow$. 
Then, $u_{n}({\bf r}) = U_{\vec {k}}e^{i(\vec {k}+{\vec q})\cdot \bf r}$ and 
$v^{*}_{n}({\bf r}) = V_{\vec {k}} e^{i(-\vec {k}+{\vec q})\cdot\bf r}$. 
\vskip 1pc

\vspace{2pc}
\hspace{2pc}
\noindent

\leftline{\large\bf 1.6. Relation between pair potential and gap parameter}

\vspace{1pc}

From Y.-J. Kim, ``Pairing constraint on the real space formalism of the theory of superconductivity". 

\vspace{1pc}

For a homogeneous system, it was shown
\begin{eqnarray}
\Delta({\bf r}-{\bf r'})=\int d{\vec k} e^{i{\vec k}\cdot({\bf r}-{\bf r'})}\Delta_{{\vec k}}.
\end{eqnarray}
But this relation is not exact because of the BCS retardation
cutoff. 

Correct relation may be obtained only after incorporating the pairing constraint
into the self-consistency equation.
It is given
\begin{eqnarray}
\Delta({\bf r}-{\bf r'})=V\sum_{{\vec k}}{\Delta_{{\vec k}}\over 2E_{{\vec k}}}tanh
{E_{{\vec k}}\over 2T}\phi_{{\vec k}}({\bf r})\phi_{-{\vec k}}({\bf r'}).
\end{eqnarray}
Comparing Eq. (50) with the BCS gap equation,  
 we also find  
\begin{eqnarray}
\Delta_{{\vec k}}=\int \phi_{{\vec k}}^{*}({\bf r})\phi^{*}_{-{\vec k}}({\bf r})
\Delta({\bf r})d{\bf r}.
\end{eqnarray}

In the presence of impurities, one finds that
\begin{eqnarray}
\Delta({\bf r})=V\sum_{n}{\Delta_{n}\over 2E_{n}}tanh
{E_{n}\over 2T}\psi_{n}({\bf r})\psi_{\bar n}({\bf r}),
\end{eqnarray}
and
\begin{eqnarray}
\Delta_{n}=\int \psi_{n}^{*}({\bf r})\psi^{*}_{\bar n}({\bf r})
\Delta({\bf r})d{\bf r}.
\end{eqnarray}
Eq. (53) was obtained first by M. Ma and P. A. Lee.

\vspace{2pc}

\hspace{2pc}
\noindent

\leftline{\large\bf 1.7. Problem in the microscopic derivation of the Ginzburg-}

\hspace{5pc} {\bf Landau theory}

\vspace{1pc}

From Y.-J. Kim, ``Pairing constraint on the real space formalism of the theory of superconductivity". 

\vspace{1pc}

Gor'kov's microscopic derivation of the Ginzburg-Landau theory is not valid. [L. P. Gor'kov,
Sov. Phys. JETP, {\bf 9}, 1364 (1959).] The problem is in
using the self-consistency equation which violates the physical constraint of the Anomalous 
Green's function. Since
the local free energy is not well defined for the Cooper pairs as for the hard-core particles,
the gradient term may not be derived microscopically. 

From the physical constraint of the Anomalous Green's function,
\begin{equation}
F({\bf r},{\bf r'}) = F({\bf r}-{\bf r'}),
\end{equation}
we obtain the revised self-consistency equation,
\begin{eqnarray}
\Delta({\bf r}) &=& VT\sum_{\omega}\int \Delta({{\bf l}})\{G^{\uparrow}_{\omega}({\bf r,{\bf l}})
G^{\downarrow}_{-\omega}({\bf r,{\bf l}})\}^{P}d{\bf l},\\
&\not=& \rm{Ginzburg-Landau\ equation}\\
&=& \rm{BCS\ gap\ equation}.
\end{eqnarray}

\vfill\eject

\hspace{2pc}

\vspace{2pc}

\leftline{\large\bf 2. Reinvestigation of impure superconductors}

\vspace{2pc}

 2.1. History of the theory of impure superconductors

 2.2. New theory of impure superconductors by Kim and Overhauser

 2.3. Strong-coupling theory of impure superconductors

 2.4. Compensation of magnetic impurity effect by radiation damage 

\hspace{2pc} or ordinary impurity

 2.5. Localization and superconductivity

\vspace{4pc}
\hspace{2pc}
\noindent

\centerline{\large\bf 2.1. History of the theory of impure superconductors} 

\vspace{2pc}

\centerline{{\bf TABLE I}. Theories of impure superconductors}
\vskip 4pt
\begin{tabular}{lll}\hline\hline
 & Ordinary impurity & Magnetic impurity\\ \hline
Anderson & $T_{c} =T_{co}$ \qquad & \\ 
AG & $T_{c} = T_{co}-{T_{co}\over \pi\omega_{D}\tau}({1\over \lambda}+{1\over 2})$  \qquad & $T_{c}=T_{co}-{\pi\over 4}{1\over \tau_{s}}$\\ 
Suhl and Matthias & $T_{c}\cong T_{co}-{T_{co}\over \lambda\omega_{D}\tau}$ \qquad & 
$T_{c}=T_{co}-{\pi\over 3.5}{1\over \tau_{s}}$\\ 
Baltensperger &  \qquad & $T_{c}=T_{co}-{\pi\over 4}{1\over \tau_{s}}$\\
Kenworthy \& ter Haar & $T_{c}\cong T_{co}-{T_{co}\over \lambda\omega_{D}\tau}$ 
    \qquad & \\
Tsuneto & $T_{c}=T_{co}$ \qquad & \\
KO & $T_{c}=T_{co}-{T_{co}\over \pi\lambda E_{F}\tau}$ \qquad & $T_{c}
=T_{co}-{0.18\pi\over \lambda\tau_{s}}$\cr
& \qquad $-$ Weak Loc. correction (Kim) \qquad & \\ \hline\hline
\end{tabular}
\vskip 8pt

\vspace{2pc}

$\bullet$ It is ironic that Kenwhorthy and ter Haar [Phys. Rev, {\bf 123}, 1181 (1961)] 
said that Abrikosov and Gor'kov's (AG) theory of impure superconductors is wrong 
because of the absence of the correction term $1\over \omega_{D}\tau$. Whereas AG 
admitted the existence of the correction term $1\over \omega_{D}\tau$ in their 
theory. 

\vspace{1pc}

{\sl Strictly speaking, in the electron-electron interaction model under consideration, this conclusion is true only to within terms of order $1/\omega_{D}\tau\sim 10^{-6}
cm/\ell$}.
(From p. 337 in ``Methods of quantum field theory in statistical physics".)

It is also amazing that ter Haar translated the AG's paper[Sov. Phys. JETP, {\bf 12}, 1243
(1961)].

\vspace{2pc}
\hspace{2pc}
\noindent

\centerline{\large\bf 2.2. New theory of impure superconductors by Kim and}
\centerline{\large\bf Overhauser} 

\vspace{1pc}

{\bf 2.2.1 Ordinary impurity case}

Recently, Kim and Overhauser (KO)$^{1}$ showed the following: 

{\bf (i)} Abrikosov and Gor'kov's (AG) theory$^{2}$ of an impure superconductor 
predicts a large decrease of $T_{c}$, proportional to $1/\omega_{D}\tau$. 
$\omega_{D}$ denotes the Debye frequnecy
and $\tau$ is the scattering time, respectively.

$\Longrightarrow$ 
\noindent
The existence of the above correction term was confirmed by Abrikosov, Gor'kov
and Dzyaloshinskii,$^{4}$  and was also shown by other workers.$^{5-7}$ 
The correction term is related with the change of electron density of states
due to the impurity scattering.
However, the correct value was shown to be $1/E_{F}\tau$.$^{1}$
Here $E_{F}$ denotes the Fermi energy.

\vspace{1pc}

{\bf (ii)} Anderson's theorem$^{3}$ is valid only to the first power in the impurity
concentration. For strongly localized states, the phonon-mediated interaction
is exponentially small.

$\Longrightarrow$ 
\noindent
It is, then, expected that weak localization correction terms occur in the
phonon-mediated interaction.

\vspace{1pc}

{\bf Table I.} Mean free path and the phonon-mediated interaction
in dirty, weak localization and strong localization limits.
Here $\ell$ and $L$ are the elastic and inelastic mean free paths and 
$\alpha$ denotes the inverse localization length.

\vspace{1pc}

\begin{tabular}{lrrr}\hline
{disorder limit } & \hspace{2pc} { dirty } & \hspace{3pc} { weak localization  \hspace{2pc}} &\hspace{1pc}  { strong localization }  \\ \hline
$\hspace{2pc}\ell$  & $ \sim 100\AA$ & $\sim 10\AA\hspace{4pc}$
                   & $\sim 1\AA$ \hspace{3.5pc} \vspace{1pc}\\ 
$\hspace{1pc}V_{mm'}$  & $V$\hspace{1.0pc} & $V[1-{2\over\pi k_{F}\ell}ln(L/\ell)] \hspace{1.0pc}(2d)$
   & $\sim exp(-\alpha L)$\hspace{2pc} \\ 
  &   &  $V[1-{3\over(k_{F}\ell)^{2}}(1-{\ell\over L})] \hspace{1.0pc}(3d)$
                           & \\ 
  &   &  $V[1-{1\over(\pi k_{F}a)^{2}}({L\over \ell}-1)] \hspace{1.0pc}(1d)$
                           & \\ \hline
\end{tabular}

\vspace{1pc}

[From Y.-J. Kim, Mod. Phys. Lett. B {\bf 10}, 555 (1996)].
\vspace{3pc}

{\bf 2.2.1 Magnetic impurity case}

For magnetic impurity effects, Kim and Overhauser (KO)$^{8}$ also proposed a BCS type
theory with different predictions: 

{\bf (i)} The initial slope of $T_{c}$
decrease depends on the superconductor and is not the universal
constant proposed by Abrikosov and Gor'kov(AG).$^{9}$ 

\begin{equation}
T_{c} =T_{co}-{0.18\pi\over \lambda\tau_{s}}.
\end{equation}

\vspace{1pc}

{\bf (ii)} The $T_{c}$ reduction
by exchange scattering is partially suppressed by potential scattering
when the overall mean free path is smaller than the coherence length.
This compensation has been confirmed in several experiments.$^{10-14}$

Note that if we impose a pairing constraint on the self-consistency equation or 
the AG's calculation, we can find KO's result.$^{15}$

\vspace{2pc}

{\bf References}
\hb
1. Y.-J. Kim and A. W. Overhauser, Phys. Rev. B {\bf 47}, 8025 (1993).\hb
2. A. A. Abrikosov and L. P. Gor'kov, Zh. Eksp. Teor. Fiz. {\bf 39}, 1781 (1961)

[Sov.  Phys. {\bf JETP 12}, 1243 (1961)].\hb
3. P. W. Anderson, J. Phys. Chem. Solids, {\bf 11}, 25 (1959).\hb
4. A. A. Abrikosov, L. P. Gor'kov,and I. Ye. Dzyaloshinskii, {\sl Methods of 

Quantum Field Theory in Statistical Physics}, (Dover Publications, New York, 

1975), p. 337.\hb
5. H. Suhl and B. T. Matthias, Phys. Rev. {\bf 114}, 977 (1959).\hb 
6. K. Nakamura, Prog. Theo. Phys. {\bf 21}, 435 (1959).\hb 
7. D. J. Kenworthy and D. etr Haar, Phys. Rev. {\bf 123}, 1181 (1961).\hb
8. Yong-Jihn Kim and A. W. Overhauser, Phys. Rev. B{\bf 49}, 15779 (1994).\hb
9. A. A. Abrikosov and L. P. Gor'kov, Sov. Phys. JETP {\bf 12}, 1243 (1961).\hb
10. M. F. Merriam, S. H. Liu, and D. P. Seraphim, Phys. Rev. {\bf 136}, A17 (1964).\hb
11. G. Boato, M. Bugo, and C. Rizzuto, Phys. Rev. {\bf 148}, 353 (1966).\hb
12. G. Boato and C. Rizzuto, (to be published), (referenced in A. J. Heeger, 

(1969), in Solid State Physics {\bf 23}, eds. F. Seitz, D. Turnbull and H. 

Ehrenreich (Academic Press, New York), p. 409)  \hb
13. W. Bauriedl and G. Heim, Z. Phys. B{\bf 26}, 29 (1977); M. Hitzfeld and G. 

Heim, Sol.   Sta. Com. {\bf 29}, 93 (1979).\hb  
14. A. Hofmann, W. Bauriedl, and P. Ziemann, Z. Phys. B{\bf 46}, 117 (1982).\hb
15. Y.-J. Kim, Mod. Phys. Lett. B {\bf 10}, 555 (1996).

\vspace{2pc}
\hspace{2pc}
\noindent

\centerline{\large\bf 2.3. Strong-coupling theory of impure superconductors }

\vspace{1pc}

Recently, Abrikosov and Gor'kov$^{1}$ argued that  the
correction term $1/\omega_{D}\tau$ in their theory disappears in the Eliashberg equation 
apart from the corrections of the order  $1/E_{F}\tau$. 
In fact, this result was first obtained by Tsuneto.$^{2}$
As a result, they admittedly showed that Gor'kov formalism is inconsistent
with the Eliashberg equation.

At this point, we may need to pause to answer the following deep 
question:
Is there a correspondence rule between strong-coupling and weak-coupling theories of impure superconductors?
The answer is yes.
It can be shown that the correspondence principle, which
relates strong-coupling and weak-coupling theories, works provided that 
Anderson's pairing condition is satisfied.

\vspace{1pc}

{\bf 2.3.1. Strong-coupling theory with Anderson's pairing} 

\vspace{1pc}

The conventional strong-coupling self-consistency equation is$^{3}$

\begin{eqnarray}
\Delta^{*}(&\omega_{n}&, {\bf r})Z(\omega_{n}) \nonumber\\
 &=& \gamma^{2}  T\sum_{n'}\lambda(\omega_{n},\omega_{n'})\int d{\bf r}_{o}
G_{N}^{\uparrow}(-\omega_{n'},{\bf r}_{o},
{\bf r})G_{N}^{\downarrow}(\omega_{n'},{\bf r}_{o},{\bf r})\Delta^{*}(\omega_{n'},{\bf r}_{o})
Z(\omega_{n'}).  
\end{eqnarray}
Equation (59) states physically that the pair potential $\Delta^{*}(\omega_{n'},{\bf r}_{o})$
launches (from the regions near ${\bf r}_{o}$) electron pairs which collaborate to generate
a pair potential $\Delta^{*}(\omega_{n},{\bf r})$ in the region near ${\bf r}$.
However, Eq. (59) misses the most important information of Anderson's pairing condition.
Whereas it was shown$^{4}$ that Anderson's pairing condition is 
derived from the physical constraint of the Anomalous Green's function, i.e.,
\begin{eqnarray}
\overline{F^{+}(\omega_{n},{\bf r},{\bf r'})}^{imp} &=& 
\overline{F^{+}(\omega_{n},{\bf r}-{\bf r'})}^{imp}, \\
\overline{\Delta^{*}(\omega_{n},{\bf r})}^{imp} &=& 
\overline{ \Delta^{*}(\omega_{n})}^{imp}.
\end{eqnarray}
Consequently, the revised self-consistency equation is
\begin{eqnarray}
\Delta^{*}(&\omega_{n}&, {\bf r})Z(\omega_{n})  = \nonumber\\
 & \gamma^{2} & T\sum_{n'}\lambda(\omega_{n},\omega_{n'})\int d{\bf r}_{o}
\{G_{N}^{\uparrow}(-\omega_{n'},{\bf r}_{o}, {\bf r})
G_{N}^{\downarrow}(\omega_{n'},{\bf r}_{o},{\bf r})\}^{P}
\Delta^{*}(\omega_{n'},{\bf r}_{o})Z(\omega_{n'}), 
\end{eqnarray}
where
$P$ denotes Anderson's pairing constraint. 

The importance of Anderson's pairing constraint was already noticed by Ma and Lee.$^{5}$
They showed that the gap parameter is given by
\begin{equation}
\Delta^{*}(\omega_{n}, m) = \int \psi_{m}({\bf r})\psi^{*}_{m}({\bf r})
\Delta^{*}(\omega_{n},{\bf r})d{\bf r}.
\end{equation}
Substitution of Eq. (63) into Eq. (62) leads to a strong-coupling gap equation
\begin{equation}
\Delta^{*}(\omega_{n}, m)Z(\omega_{n}) = T
\sum_{n'}\lambda(\omega_{n},\omega_{n'})
 \sum_{m'}V_{mm'}{\Delta^{*}(\omega_{n'},m')Z(\omega_{n'})\over [\omega_{n'}Z(\omega_{n'})]^{2}
+\epsilon_{m'}^{2}},
\end{equation}
where
\begin{equation}
V_{mm'} = \gamma^{2}\int |\psi_{m}({\bf r})|^{2}
 |\psi_{m'}({\bf r})|^{2}d{\bf r}.
\end{equation}

\vspace{1pc}

{\bf 2.3.2. Comparison with Tsuneto's theory}

Tsuneto$^{2}$ obtained the gap equation
\begin{eqnarray}
\Sigma_{2}(\omega) = {i\over (2\pi)^{3}p_{o}}\int dq\int d\epsilon \int d\omega '
{qD(q,\omega-\omega')\eta(\omega')\Sigma_{2}(\omega')\over \epsilon^{2} - 
\eta^{2}(\omega')\omega'^{2}},
\end{eqnarray}
where $\eta=1 +{1\over 2\tau|\omega|}$, and $\tau$ is the collision time.

Comparing Eqs. (64) and (66), we find that Tsuneto's result misses
the most important factor $V_{mm'}$, which gives the change of the
phonon-mediated interaction due to impurities. 
This factor is exponentially small for the localized
states. 
In the weak localization limit, it was shown that$^{6}$ 
\begin{eqnarray}
V_{mm'}^{3d} &\cong& 
-V[1-{1\over (k_{F}\ell)^{2}}(1-{\ell\over L})],\nonumber \\
V_{mm'}^{2d} &\cong&  
 -V[1-{2\over \pi k_{F}\ell}ln(L/\ell)],\nonumber \\
V_{mm'}^{1d} &\cong&  
 -V[1-{1\over (\pi k_{F}a)^{2}}(L/\ell-1)],
\end{eqnarray}
where $a$ is the radius of the wire.

\vspace{1pc}

{\bf References}
\hb
1. A. A. Abrikosov and L. P. Gor'kov, Phys. Rev. B {\bf 49}, 12337 (1994).\hb
2. T. Tsuneto, Prog. Theo. Phys. {\bf 28}, 857 (1962).\hb 
3. G. Eilenberger and V. Ambegaokar, Phys. Rev. {\bf 158}, 332 (1967).\hb
4. Y.-J. Kim, Mod. Phys. Lett. B {\bf 10}, 353 (1996).\hb 
5. M. Ma and P. A. Lee, Phys. Rev. B{\bf 32}, 5658 (1985).\hb
6. Y.-J. Kim, Mod. Phys. Lett. B {\bf 10}, 555 (1996).

\vspace{2pc}
\hspace{2pc}
\noindent

\centerline{\large\bf 2.4. Compensation of magnetic impurity effect by radiation}
\centerline{ \large\bf damage or ordinary impurity }

\vspace{1pc}

Compensation of the reduction of $T_{c}$ caused by magnetic impurities
has been observed as a consequence of radiation damage or ordinary impurity.
The recent theory by Kim and Overhauser (KO) 
gives a good fitting to the experimental data.

Note also that Gor'kov's formalism with the pairing constraint derived from  
the Anomalous Green's function leads to KO theory.

\vspace{1pc}

$\bullet$ Compensation by radiation damage

\vspace{1pc}

{\bf Fig. 1.} Superconducting transition temperature $T_{c}$ of In (open 
symbols) and In-Mn (closed symbols) vs Ar fluence. Data are due to
Hofmann, Bauriedl, and Ziemann, Z. Phys. B {\bf 46}, 117 (1982).
 $1/ \tau_{s}$ was adjusted
in the theoretical curve (lower curve) so that $T_{co}=1.15K$ without 
irradiation. 
[From Park, Lee, and Kim, preprint].

\vspace{2in}
\vspace{1pc}

$\bullet$ Compensation by ordinary impurity

\vspace{1pc}

{\bf Fig. 2} Compensation of $T_{c}$ in In-Mn-Pb vs Pb concentration for
5-ppm Mn. Data are due to Merriam, Liu, and Seraphim, Phys. Rev. {\bf 136}, A17 (1964).
[From, Kim and Overhauser, Phys. Rev. B {\bf 49}, 15779 (1994)].

\vspace{1pc}

\vspace{2in}

\hspace{2pc}
\noindent

\centerline{\large\bf 2.5. Localization and Superconductivity}

\vspace{1pc}
For thin films, the empirical formula is given$^{1}$
\begin{equation}
{T_{co}-T_{c}\over T_{co}} \propto {1\over k_{F}\ell} \propto R_{\Box},
\end{equation}
where $T_{co}$ is the unperturbed value of $T_{c}$ and $R_{\Box}$ is the sheet resistance.
On the other hand, bulk materials show$^{2,3}$ 
\begin{equation}
{T_{co}-T_{c}\over T_{co}} \propto {1\over (k_{F}\ell)^{2}}.
\end{equation}
Notice that these results are obtained if we substitute the matrix elements in Table I 
into the (strong-coupling or weak-coupling) gap equation.

\vspace{1pc}

{\bf Table I.} Mean free path and phonon-mediated interaction
in dirty, weak localization and strong localization limits.
Here $\ell$ and $L$ are the elastic and inelastic mean free paths and 
$\alpha$ denotes the inverse localization length.

\vspace{1pc}

\begin{tabular}{lrrr}\hline
{disorder limit } & \hspace{2pc} { dirty } & \hspace{3pc} { weak localization  \hspace{2pc}} &\hspace{1pc}  { strong localization }  \\ \hline
$\hspace{2pc}\ell$  & $ \sim 100\AA$ & $\sim 10\AA\hspace{4pc}$
                   & $\sim 1\AA$ \hspace{3.5pc} \vspace{1pc}\\ 
$\hspace{1pc}V_{mm'}$  & $V$\hspace{1.0pc} & $V[1-{2\over\pi k_{F}\ell}ln(L/\ell)] \hspace{1.0pc}(2d)$
   & $\sim exp(-\alpha L)$\hspace{2pc} \\ 
  &   &  $V[1-{3\over(k_{F}\ell)^{2}}(1-{\ell\over L})] \hspace{1.0pc}(3d)$
                           & \\ 
  &   &  $V[1-{1\over(\pi k_{F}a)^{2}}({L\over \ell}-1)] \hspace{1.0pc}(1d)$
                           & \\ \hline
\end{tabular}

\vspace{1pc}

[From Y.-J. Kim, Mod. Phys. Lett. B {\bf 10}, 555 (1996)].
\vspace{2pc}

Notice that these results are obtained if we substitute the matrix elements in Table I 
into the (strong-coupling or weak-coupling) gap equation.

\vspace{2pc}

$\bullet$ {\bf 3 dimension}

{\bf Fig. 1} Calculated $T_{c}$ versus resistivity $\rho$ for 3-dimensional $\rm Nb_{3}Ge$ (dotted line)
and $\rm V_{3}Si$ (solid line). Experimental data are from J. M. Rowell and R. C. Dynes, unpublished.
[From Kim and Chang, preprint (1997)].

\vfill\eject

$\bullet$ {\bf 2 dimension}

{\bf Fig. 1} Calculated $T_{c}$ versus sheet resistance $R_{\Box}$ for a-MoGe (solid line) and Mo-C 
(dotted line) thin films. Experimental data for a-MoGe and Mo-C are from ref. 4 and 5, respectively.
[From Kim and Chang, preprint (1997)].

\vspace{5in}

{\bf References}

\vspace{1pc}\hb
1. B. I. Belevtsev, Sov. Phys. Usp. {\bf 33}, 36 (1990).\hb 
2. A. F. Fiory and A. F. Hebard, Phys. Rev. Lett. {\bf 52}, 2057 (1984).\hb 
3. S. J. Bending, M. R. Beasley, and C. C. Tsuei, Phys. Rev. B {\bf 30}, 6342 (1984).\hb
4. J. M. Graybeal, M. R. Beasley, and R. L. Greene, Physica B {\bf 29}, 731 (1984).\hb
5. S. J. Lee and J. B. Ketterson, Phys. Rev. Lett. {\bf 64}, 3078 (1990).\hb

\vfill\eject

\hspace{2pc}

\leftline{\large\bf 3. Study of high Tc superconductors}

\vspace{2pc}

 3.1. Impurity scattering in a d-wave superconductor

 3.2. Impurity doping effect in high Tc  superconductors

 3.3. On the mechanism of high Tc superconductors

 3.4. Search for new high Tc superconductors 

\vspace{3pc}

\centerline{\large\bf 3.1. Impurity scattering in a d-wave superconductor} 

\vspace{2pc}

For a d-wave superconductor, the pairing interaction $V_{{\vec k}, {\vec k}'}$
for the plane states is taken to be$^{1}$
\begin{equation}
V_{{\vec k},{\vec k}'} = \int e^{i({\vec k}-{\vec k}')\cdot {\vec r}}V({\br})d^{3}r = - 5V_{2}
 \ {1\over 2}[3({\hat k}\cdot{\hat k'})^{2}-1],
\end{equation}
where $\hat k$ is the unit vector parallel to $\vec k$.
Substituting Eq. (70) into the BCS gap equation, one finds
\begin{equation}
\Delta_{\vec k} = 5V_{2}\sum_{\vk'}
 {1\over 2}[3({\hat k}\cdot{\hat k'})^{2}-1]
{\Delta_{\vk'} \over 2E_{\vk'}}
tanh{E_{\vk'}\over 2T}, 
\end{equation}
where
\begin{equation}
E_{\vec k'} = \sqrt{\epsilon_{\vec k'}^{2} + |\Delta_{\vec k'}|^{2}},
\end{equation}
and $\epsilon_{\vec k}$ is the electron energy.
Among the possible solutions, we consider
\begin{equation}
\Delta_{\vec k} = \Delta_{o}({\hat {k}_{x}}^{2} - {\hat {k}_{y}}^{2}).
\end{equation}
This solution has the same symmetry property as $d_{x^{2}-y^{2}}=\Delta_{o}
(cos k_{x}-cos k_{y})$ which is believed to describe the gap structure
of the cuprate high $T_{c}$ superconductors.

\vspace{1pc}

{\bf  Non-Magnetic Impurity Effect}

\vspace{1pc}

In the presence of impurities, the scattered states $\psi_{n}$ may be expanded
in terms of plane waves, such as$^{2}$
\begin{equation}
\psi_{n}=\sum_{\vec k}e^{i{\vec k}\cdot{\bf r}}<{\vec k}|n>.
\end{equation}
Now the pairing interaction $V_{nn'}$ between scattered basis pairs
$(\psi_{n}, \psi_{\bar n})$ and $(\psi_{n'}, \psi_{\bar n'})$ is given by
\begin{equation}
V_{nn'}=\int\int d{\bf r}_{1}d{\bf r}_{2}\psi_{n'}^{*}({\bf r}_{1})
\psi_{\bar n'}^{*}({\bf r}_{2})V(|{\bf r}_{1}-{\bf r}_{2}|)
\psi_{\bar n}({\bf r}_{2}) \psi_{n}({\bf r}_{1}).
\end{equation}
Here $\psi_{\bar n}$ denotes the time-reversed state of $\psi_{n}$.
From Eqs. (70), (74) and (75) we can calculate $V_{nn'}$.

The pairing interaction is reduced: 
\begin{equation}
V_{{\vec k},{\vec k}'} = - 5V_{2}\ {1\over 2}[3({\hat k}\cdot{\hat k'})^{2}-1]
 [1 + {3.5\xi_{o}\over 4\ell}]^{-2}.
\end{equation}
Notice that in dilute limit the reduction is proportional
to the ratio of the average correlation length to the mean free
path, $\xi_{o}/\ell$ and the pairing interaction decreases
linearly with the impurity concentration.
The $T_{c}$ equation is now,
\begin{equation}
T_{c} = 1.13\epsilon_{c}e^{-1/ N_{o}V_{2}[1+{3.5\xi_{o}\over 4\ell}]^{-2}}. 
\end{equation}
Figure 1 shows $T_{c}$ versus $1/\tau$ for $T_{co}=40K\ \rm{and}\ 80K$ 
respectively. $T_{co}$ denotes the transition temperature without impurities.
We used $\epsilon_{c}=500K$. For a metal with $v_{F}=2\times 10^{7} cm/sec$,
the superconductivity is completely suppressed when the mean free paths are
about $1000 \AA$ and $350\AA$ for $T_{co}=40K\ \rm {and}\ 80K$, respectively.  

\vspace{2.5in}
\vskip 1pt\hb
{\bf Fig. 1} Variation of $T_{c}$ with impurity concentration (measured 
in terms of the scattering rate, $1\over \tau$) for $T_{co}=40K$ and
$80K$, respectively. The cutoff energy $\epsilon_{c}$ is $500K$.

\vskip 1pc

{\bf Discussions}

\vskip 1pc

In high $T_{c}$ superconductors, the impurity doping 
and ion-beam induced damage$^{3}$ suppress strongly $T_{c}$. 
But the $T_{c}$ reduction 
is not fast enough to be explained by this study. 
The experimental data show that $T_{c}$ reduction is closely
related with the proximity to a metal-insulator transition caused
by the impurity doping and the ion-beam-induced damage.$^{3-5}$
It seems that the local fluctuations of the gap parameter near the
impurities may decrease the effect of impurities in the dirty limit. 
[From Park, Lee, and Kim, Mod. Phys. Lett. B {\bf 11}, 719 (1997)].
\vskip 1pc

 {\bf References} 
\hb
1. P. W. Anderson and P. Morel, Phys. Rev. {\bf 123}, 1911 (1961).\hb
2. P. W. Anderson, J. Phys. Chem. Solids {\bf 11}, 26 (1959).\hb
3. Y. Li, G. Xiong, and Z. Gan, Physica C {\bf 199}, 269 (1992).\hb
4. G. Xiao, M. Z. Cieplak, J. Q. Xiao, and C. L. Chien, Phys. Rev. B {\bf 42},
    8752 (1990).\hb
5. V. P. S. Awanda, S. K. Agarwal, M. P. Das, and A. V. Narlikar, J. Phys.:
    Condens. 

Matter {\bf 4}, 4971 (1992).\hb

\vspace{2pc}
\hspace{2pc}
\noindent

\centerline{\large\bf 3.2. Impurity doping effect in high Tc superconductors} 

\vspace{2pc}

It has been observed that impurity doping and/or ion-beam-induced
damage in high $T_{c}$ superconductors 
cause a metal-insulator transition and thereby suppress 
the critical temperature.
Based on my theory of weak localization effect
on superconductivity, I examined the variation of
$T_{c}$ with increasing of impurity concentration $\rm (x)$ in
$\rm{La_{1.85}Sr_{0.15}Cu_{1-x}A_{x}O_{4}}$ systems, 
where A $=$ Fe, Co, Ni, Zn, or Ga.
The doping impurity decreases the pair-scattering matrix elements,
 such as
 $V_{nn'}=-V[1-{2\over \pi k_{F}\ell} ln(L/\ell)]$,
where $L$ and $\ell$ are the inelastic and elastic mean
free paths, respectively. Using the mean free path $\ell$ determined from 
resistivity data,  we find good agreements between 
calculated values for $T_{c}$ and experimental data except Ni-doped
case. [See Kim and Chang, preprint (1997)].

\vspace{1pc}

\noindent
{\bf Fig. 1} Variation of $T_{c}$ with dopant concentration 
for $\rm La_{1.85}Sr_{0.15}Cu_{1-x}Ga_{x}O_{4}$. 
Experimental data are from [Xiao, Cieplak, Xiao, and Chien, PR B {\bf 42},
8752 (1990)]. 

\noindent

\hspace{2pc}

\vspace{1.8in}

{\bf Fig. 2} Variation of $T_{c}$ with dopant concentration 
for $\rm La_{1.85}Sr_{0.15}Cu_{1-x}Fe_{x}O_{4}$. 

\vspace{1.8in}

\hspace{2pc}
\noindent

\centerline{\large\bf 3.3. On the mechanism of high Tc superconductors} 

\vspace{2pc}

In Ni-doped case, $T_{c}$ suppression is much slower than expected.
This implys that Ni may enhance the pairing interaction in high $T_{c}$ 
superconductors such as LSCO, YBCO(123), and YBCO(124). 
It is also interesting that Ni impurity in YBCO(123) acts as an unpaired
spin of $S={1\over 2}$ rather than $S=1$ expected for $\rm Ni^{2+}$.$^{17}$
Further microscopic study on Ni-doped samples may give a clue to understanding 
the mechanism of high $T_{c}$ superconductors.
In particular, I am calculating the electronic structure of Ni-doped samples 
by the exact diagonalization of finite-size clusters.

\vskip 2pc

\vfill\eject
\noindent
{\bf Fig. 1} Variation of $T_{c}$ with Ni concentration
for $\rm La_{1.85}Sr_{0.15}Cu_{1-x}Ni_{x}O_{4}$.

\vspace{1.5in}

\hspace{2pc}
\noindent

\centerline{\large\bf 3.4. Search for new high Tc superconductors} 

\vspace{1pc}

$\rm La_{2}NiO_{4}$ is a Mott-Hubbard insulator consisting of 
antiferromagnetic $\rm NiO_{2}$ planes as $\rm La_{2}CuO_{4}$.$^{1}$
However, $\rm La_{2-x}Sr_{x}NiO_{4}$ remains nonmetallic until $x\geq 0.8$
because the holes doped into the $\rm NiO_{2}$ planes tend to order
themselves in periodically spaced stripes.$^{2-4}$
There is also evidence for related stripe correlations in hole-doped
$\rm La_{2}CuO_{4}$.$^{5-7}$

The stripe order seems to localize the holes.
So we need to study how to induce metallic phase and concomitant 
superconductivity in $\rm La_{2-x}Sr_{x}NiO_{4}$.
One possibility is to dope a large amount of Cu into the $\rm NiO_{2}$ planes
in order to increase the mobility of the holes.
{\sl If Cu and Ni order in the planes, $T_{c}$ may be very high}.
Another possibility is to substitute O by N, F, or other elements, in order
to increase the hybridization between 3d and 2p orbitals.
Then, the system may become metallic and superconducting.

\vskip 1pc
\vskip 0.3pc
{\bf References}
\hb
1. H. Eisaki et al., Phys. Rev. B 45, 12513 (1992).\hb
2. C. H. Chen, S.-W. Cheong, and A. S. Cooper, Phys. Rev. Lett. 71, 2461 
    (1993).\hb
3. J. M. Tranquada, D. J. Buttrey, V. Sachan, and J. E. Lorenzo, Phys. Rev. Lett.  

73, 1003 (1994).\hb
4. V. Sachan et al., Phys. Rev. B 51, 12742 (1995).\hb
5. S.-W. Cheong et al., Phys. Rev. Lett. 67, 1791 (1991).\hb
6. T. E. Mason, G. Aeppli, and H. A. Mook, Phys. Rev. Lett. 68, 1414 (1992).\hb
7. J. M. Tranquada et al., Nature 375, 561 (1995).
\vfill\eject
\hspace{2pc}

\leftline{\large\bf 4. New formalism for inhomogeneous superconductors}

\vspace{2pc}

 4.1. Generalization of the BCS theory

 4.2. Generalization of the Ginzburg-Landau theory

\vspace{2pc}

This chapter is still in progress.

\vspace{1pc}

\centerline{\large\bf 4.1. Generalization of the BCS theory} 

\vspace{1pc}

We may distinguish the impurity problem and the junction 
or the vortex problem. When we consider the impurity problem using
Anderson's approach, the local fluctuation of the gap parameter
is disregarded. However, since the latter two problems are related to
the macroscopic or mesoscopic inhomogeneity, we should allow
the gap parameter to vary as a function of the position.
Accordingly, we need to generalize the BCS theory to tackle
the macroscopically or mesoscopically inhomogeneous systems.

A key idea is to introduce the position-dependent Cooper-pair size
into a BCS type wavefunction.

\vspace{2pc}

\centerline{\large\bf 4.2. Generalization of the Ginzburg-Landau theory} 

\vspace{1pc}

The above study may lead to an extention of the Ginzburg-Landau theory
on the macroscopic level. Because the Ginzburg-Landau theory
appeared before the pairing theory of BCS, it lacks the effect of the Cooper 
pair-size. 
 The Ginzburg-Landau coherence
length $\xi_{GL}$ is the characteristic length for variation of the 
order parameter. This $\xi_{GL}$ is not the same length as the BCS coherence 
length $\xi_{o}$. Note that the local free energy density is not well defined in superconductors 
because of the Cooper-pair size.  

It is very important to incorporate the information of the pairing correlation
into the traditional Ginzburg-Landau theory.

\vfill\eject

\hspace{2pc}

\vspace{2pc}

\leftline{\large\bf 5. Future directions}

\vspace{2pc}

\hspace{1pc} 5.1. Impure superconductors

\hspace{1pc} 5.2. Localization and Superconductivity

\hspace{1pc} 5.3. Proximity effect

\hspace{1pc} 5.4. Andreev reflection

\hspace{1pc} 5.5. Josephson effect

\hspace{1pc} 5.6. Magnetic field effect

\hspace{1pc} 5.7. Type II superconductors

\hspace{1pc} 5.8. Vortex problem

\hspace{1pc} 5.9. Mesoscopic superconductivity

\hspace{1pc} 5.10. Non-equilibrium superconductivity

\hspace{1pc} 5.11. Granular superconductors 

\hspace{1pc} 5.12. High $T_{c}$ superconductors

\hspace{2pc}

\vspace{4pc}

\centerline{\large\bf 5.1.  Impure superconductors}

\vspace{1pc}

{\bf Ordinary impurity case}

\vspace{1pc}

Anderson's approach can be used to restudy the following the 
topics:

1. Thermodynamic properties,

2. Electrodynamics,

3. Coherence effects,

4. Response functions,

5. Strong-coupling theory using the realistic phonon model.

Superconducting behavior very near an impurity may not be understood
by Anderson's approach. We need a more general formalism which can
take into account the variation of the gap parameter near the impurity.

\vspace{2pc}

{\bf Magnetic impurity case}

\vspace{1pc}

KO theory may used to restudy the following the 
topics:

1. Thermodynamic properties,

2. Electrodynamics,

3. Coherence effects,

4. Response functions.

It is clear that compensation of the magnetic impurity effect by 
radiation damage or ordinary impurity should be subjected to further 
experimental study.

\vspace{2pc}

\centerline{\large\bf 5.2.  Localization and Superconductivity}

\vspace{1pc}

In the weak localization limit, I showed that 
\begin{eqnarray}
V_{mm'}^{3d} &\cong& 
-V[1-{1\over (k_{F}\ell)^{2}}(1-{\ell\over L})],\nonumber \\
V_{mm'}^{2d} &\cong&  
 -V[1-{2\over \pi k_{F}\ell}ln(L/\ell)],\nonumber \\
V_{mm'}^{1d} &\cong&  
 -V[1-{1\over (\pi k_{F}a)^{2}}(L/\ell-1)],\nonumber
\end{eqnarray}
where $a$ is the radius of the wire.

\vspace{1pc}

Using the above matrix elements, we may study the following problems:
 
1. Thermodynamic properties,

2. Electrodynamics,

3. Coherence effects,

4. Response functions,

5. Effect of spin-orbit scattering

6. Strong-coupling theory.

The so-called superconductor-insulator transition may
be also understood. It seems that the superconductor-insulator
transition is not a sharp phase transition but a crossover
phenomena from quasi-wto dimensional to two-dimensional.
Note that the critical sheet resistance for the suppression
of superconductivity in this films is not a universal
constant, but a sample-dependent quantity. 

\vspace{2pc}

\centerline{\large\bf 5.3.  Proximity Effect}

\vspace{1pc}

Note that Gor'kov formalism with a pairing constraint leads to the revised
self-consistency equation which is nothing but another form of the
BCS gap equation.$^{1}$ Accordingly, both the revised gap equation and
the BCS gap equation are useless in describing the proximity effect.
It is understandable that proximity effect is a long-standing unsolved 
problem.  We need a new formalism to determine how fast the Cooper-pair 
size is changing in the normal region. 

Note that the proximity effect in mesoscopic superconducting junctions 
shows anomalous behaviors.$^{2,3}$ 

{\bf References}

1. Y.-J. Kim, Mod. Phys. Lett. B {\bf 10}, 555 (1996).

2. T. M. Klapwijk, Physica B {\bf 197}, 481 (1994).

3. C. J. Lambert and R. Raimondi, preprint (1997).

\vspace{2pc}

\centerline{\large\bf 5.4.  Andreev reflection}

\vspace{1pc}

Recently, it has been realized that the distinction between
the proximity effect and Andreev reflection is artificial.$^{1}$
Consequently, the conventional theory of Andreev reflection is
not complete. A unfied theory of the proximity effect and
Andreev reflection is required. 

{\bf References}

1. C. J. Lambert and R. Raimondi, preprint (1997).

\vspace{2pc}

\centerline{\large\bf 5.5. Josephson Effect}

\vspace{1pc}

Supercurrents are found in SIS, SNS, and S-semi-S structures. 
The thickness dependence of the supercurrents are not well
understood. In particular, the recent mesoscopic S-semi-S junctions 
show anomalous behavior of the supercurrents.$^{1}$
If a unified theory of the proximity effect
and Andreev reflection is constructed, the theory may shed light on 
this problem.

Note also the sign problem in the pair-quasi-particle interference
term.$^{2}$

{\bf References}

1. T. M. Klapwijk, Physica B {\bf 197}, 481 (1994).

2. M. Tinkham, {\sl Introduction to superconductivity}, McGraw-Hill, (1975), 

\hspace{1pc} p. 195. 

\vspace{2pc}

\centerline{\large\bf 5.6. Magnetic field effect}

\vspace{1pc}

KO$^{1}$ theory may be useful in determining the magnetic field effect on 
superconductors.
Notice the discrepancies in existing thoeries.

\vspace{1pc}

{\bf References}

1. Y.-J. Kim and A. W. Overhauser, Phys. Rev. B {\bf 49}, 15799 (1994).

\hspace{2pc}

\vspace{4pc}

From B. Y. Tong, Phys. Rev. {\bf 130}, 1322 (1963).

\vspace{2in}

\hspace{2pc}

From M. Rasolt and Z. Tesanovic, Rev. Mod. Phys. {\bf 64}, 709 (1992).

\vspace{2in}

The reentrant superconducting state in very high magnetic field
may be an artifact caused by the pairing problem in Gor'kov's formalism.

\vspace{4pc}

\centerline{\large\bf 5.7. Type II superconductors}

\vspace{1pc}

The magnetic behavior of Type II superconductors may need to be reexamined.
$H_{c1}$, $H_{c2}$,  vortex, and flux pinning, creep, and flow are 
aprticularly interesting.
We need a new formalism to tackle these problems.

\vspace{2pc}

\centerline{\large\bf 5.8. Vortex problem}

\vspace{1pc}

Recent STM experiments show that the microscopic vortex structure  
is very complicated. The conventional Green's function theory is not
applicable to this problem. We had better solve one vortex problem 
using a new microscopic formalism.

\vspace{2in}

\centerline{\large\bf 5.9. Mesoscopic superconductivity}

\vspace{1pc}

Recently, much attention has been paid to this topic.$^{1,2,3}$
It is clear that our understanding of inhomogeneous superconductors
is in its infancy.

\vspace{2in}

{\bf References}

1. T. M. Klapwijk, Physica B {\bf 197}, 481 (1994).

2. M. Tinkham, Physica B {\bf 204}, 176 (1995).

3. C. J. Lambert and R. Raimondi, preprint (1997).

\vspace{2pc}

\centerline{\large\bf 5.10. Non-equilibrium superconductivity}

\vspace{1pc}

Non-equilibrium superconductivity is similar to inhomogeneous 
superconductors in that the Cooper-pair size may vary as a function
of the position. 

\vspace{2pc}

\centerline{\large\bf 5.11. Granular superconductors}

\vspace{1pc}

Granular superconductors are related to the macroscopic or mesoscopic
inhomogeneity. 

\vspace{2pc}

\centerline{\large\bf 5.12. High Tc superconductors}

\vspace{1pc}

Since high Tc superconductors are strongly correlated, 
both normal and superconducting properties are significantly 
influenced by the correlation effect. We need to know how to take into 
account properly this effect.  
For an example, the experiments clearly show that impurity potential is
strongly renormalized by correlation.

\end{document}